\documentclass[12pt]{iopart}

\usepackage{graphicx}

\begin{document}

\paper{Directed motion generated by heat bath nonlinearly driven by
external noise}

\author{J. Ray Chaudhuri$^{1*}$, 
D. Barik$^{2\dagger}$\footnote{Present address:
Department of Biology, Virginia Polytechnic Institute
and State University, Blacksburg, VA 24061-0406, USA}
and S. K. Banik$^{3\ddagger}$}

\address{$^1$Department of Physics, Katwa
College, Katwa, Burdwan 713130, West Bengal, India}
\address{$^2$Indian Association for the Cultivation of Science,
Jadavpur, Kolkata 700032, India}
\address{$^3$Department of Physics, Virginia Polytechnic Institute
and State University, Blacksburg, VA 24061-0435, USA}

\eads{\mailto{$^*$jprc$_-$8@yahoo.com},
\mailto{$^\dagger$dbarik@vt.edu},
\mailto{$^\ddagger$skbanik@phys.vt.edu}}

\begin{abstract}
Based on the system heat bath approach where the bath is nonlinearly
modulated by an external Gaussian random force, we propose a new
microscopic model to study directed motion in the overdamped limit
for a nonequilibrium open system. Making use of the coupling between
the heat bath and the external modulation as a small perturbation we
construct a Langevin equation with multiplicative noise and space
dependent dissipation and the corresponding
Fokker-Planck-Smoluchowski equation in the overdamped limit. We
examine the thermodynamic consistency condition and explore the
possibility of observing a phase induced current as a consequence of
state dependent diffusion and, necessarily, nonlinear driving of the
heat bath by the external noise.
\end{abstract}


\submitto{\JPA}

\maketitle

In recent times the phenomena of noise induced transport under
nonequilibrium condition have gained wide interdisciplinary interest
where the interplay of fluctuations and nonlinearity of the system
plays an important role \cite{Misc,RMP,Reimann,Astumian,Linke}.
Exploitation of the nonequilibrium fluctuations present in the
medium helps to generate phase induced directed motion of the
Brownian particle. Presence of spatial anisotropy in the potential
together with nonequilibrium perturbations enables one to extract
the useful work from random fluctuations without violating the
second law of thermodynamics \cite{Reimann}. This lead to its wide
applicability in explaining the dynamics of molecular motors
\cite{RMP,Howard}, directed transport in photovoltaic and
photoreflective materials \cite{Glass}, and the efficiency of tiny
molecular machine in a highly stochastic environment
\cite{Linke,Astumian,Linke2}, realization of ratchet effect in cold
atom \cite{Renzoni}, construction of artificial molecular rotors
that produce controlled directional motion mimicking molecular motor
protein \cite{Balzani}. In some special cases one can generate
directed motion even in a symmetric potential due to state dependent
diffusion. For such systems the state dependent diffusion
coefficient, $D(q)$, felt by the Brownian particle could arise
either due to space dependent friction or presence of local hotspots
\cite{RMP,Reimann,Landauer,Porto}.

To the best of our knowledge, in almost all the above mentioned
cases the corresponding Langevin equation is either written
phenomenologically or is constructed from a microscopic system heat
bath Hamiltonian model where the associated heat bath is in thermal
equilibrium. To generate directed motion one then applies an
external perturbation, time periodic force or correlated random
force, to break the symmetry of the force field as equilibrium
thermal fluctuations are unable to create spontaneous symmetry
breaking. Symmetry breaking can be also achieved by considering a
nonlinear coupling between the Brownian particle and thermal heat
bath thereby generating a multiplicative noise term in the resulting
Langevin equation which in turn gives a position dependent diffusion
term effectively creating a phase induced bias in the dynamics. In
the present paper, we propose a system heat bath model where the
heat bath is weakly modulated by an external noise. Although the
microscopic model we present here has a close kinship to our earlier
approach \cite{JRC} to study escape from a metastable state within
the context of external noise modulated heat bath, the present
formalism differs from our earlier model in the following way. The
heat bath-external noise coupling is considered to be nonlinear and
in addition to that the system is also nonlinearly coupled with the
heat bath thereby resulting in a nonlinear multiplicative Langevin
equation and the corresponding Fokker-Planck-Smoluchowski equation
with space dependent diffusion. We then explore the
possibility of observing directed transport as a result of phase
difference between the coupling function and the periodic potential
in which the Brownian particle is moving. Our theoretical
model can be tested experimentally to study the directional motion
of artificial chemical rotors in photoactive solvent \cite{Balzani}.
To observe the effect of external stochastic modulation one can
carry out the experiment in a photochemically active solvent (the
heat bath) where the solvent is under the influence of external
monochromatic light with fluctuating intensity of a wavelength which
is absorbed solely by the solvent molecules. As a result of it the
modulated solvent heats up due to the conversion of light energy
into heat energy by radiationless relaxation process and an effective
temperature like quantity develops due to the constant input of
energy. Since the fluctuations in the light intensity result in the
polarization of the solvent molecules, the effective reaction field
around the reactants gets modified \cite{Horsthemke}.

To start with we consider a classical particle of unit mass
bilinearly coupled to a heat bath consisting of $N$ mass weighted
harmonic oscillators characterized by the frequency set
${\omega_j}$. In addition to that the heat bath is nonlinearly
driven by an external noise $\epsilon(t)$. The Hamiltonian for the
composite system is
\begin{equation}\label{1}
H=H_S + H_B + H_{SB} + H_{int},
\end{equation}


\noindent where $H_S=(p^2/2)+V(q)$, is the system's Hamiltonian with
$q$ and $p$ being the coordinate and momentum of the system
particle, respectively, and $V(q)$ is the potential energy function
of the system. $H_B+H_{SB}=\sum_{j=1}^N
[(p_j^2/2)+(\omega_j^2/2)\{x_j-c_jf(q)\}^2 ]$ where $\{x_j, p_j\}$
are the variables for the $j$-th bath oscillator. The system-heat
bath interaction is given by the coupling term $c_j \omega_j f(q)$
with $c_j$ being the coupling strength. We consider the
interaction, $H_{int}=\sum_{j=1}^N \kappa_j g(x_j)
\epsilon(t)$, between the heat bath and the external noise $\epsilon
(t)$ where $\kappa_j$ denotes the strength of the interaction and
$g(x_j)$ is an arbitrary analytic function of the bath variables, in
general nonlinear. This type of interaction makes the bath variables
explicitly time dependent. A large class of phenomenologically modeled 
stochastic differential
equation may be obtained from a microscopic Hamiltonian
for particular choice of coupling function $g(x_j)$. 
In what follows we have chosen
$g(x_j)= x_j^2/2$, which makes the spring constants of the bath oscillators
time-dependent. The external noise is stationary, Gaussian with
the properties $\langle \epsilon (t) \rangle_e = 0$ and $\langle
\epsilon(t) \epsilon(t')\rangle_e = 2D \delta (t-t')$, where $D$ is
the strength of the external noise and $\langle \cdot \rangle_e$
implies averaging over the external noise processes. From
Eq.(\ref{1}) we have the dynamical equations for the system and bath
variable
\begin{eqnarray}
\ddot{q} (t) = -V'(q(t)) + f'(q(t)) 
\sum_{j}c_j \omega_j^2\{x_j (t) - c_j f(q(t))\}, \label{5} \\
\ddot{x}_j (t) + \{\omega_j^2 + \kappa_j \epsilon(t) \} x_j (t) = c_j
\omega_j^2 f(q(t)), \label{6}
\end{eqnarray}

\noindent where we have used $g(x_j)=x_j^2/2$. To solve Eq.(\ref{6})
for $x_j$ we assume a solution of the form
\begin{equation}\label{7}
x_j(t) = x_j^0 (t) + \kappa_j x_j^1 (t),
\end{equation}

\noindent where $x_j^0 (t)$ is the solution of the unperturbed equation
of motion
\begin{equation}\label{8}
\ddot{x}_j^0 (t)+ \omega_j^2 x_j^0 (t) = c_j \omega_j^2 f(q(t)).
\end{equation}

\noindent The physical situation that has been addressed here is the
following, we consider that at $t=0$, the heat bath is in thermal
equilibrium in the absence of the external noise $\epsilon(t)$. At
$t=0_+$ the external noise agency is switched on and the heat bath
is modulated by $\epsilon(t)$ \cite{JRC}. Then $x_j^1 (t)$ must satisfy
the equation
\begin{equation}\label{9}
\ddot{x}_j^1 (t) + \omega_j^2 x_j^1 (t) = - x_j^0 (t) \epsilon(t),
\end{equation}

\noindent with the initial conditions $x_j^1(0) = p_j^1(0) = 0$. The
solution of Eq.(\ref{9}) is given by
\begin{equation}\label{11}
x_j^1(t) = -\frac{1}{\omega_j}\int_0^t dt'
\sin\omega_j(t-t')x_j^0(t')\epsilon(t').
\end{equation}


\noindent
The formal solution of Eq.(\ref{8}) is given by
\begin{equation}
\label{neq1}
x_j^0 (t) = x_j^0 (0) \cos \omega_j t + \frac{p_j^0 (0)}{\omega_j}
\sin \omega_j t + c_j \omega_j \int_0^t dt' \sin \omega_j (t-t')
f (q(t'))
\end{equation}

\noindent
where $x_j^0 (0)$ and $p_j^0 (0)$ are the initial position and momentum,
respectively, of the $j$th oscillator.
Now using this solution in
Eq.(\ref{11}) we have, after an integration by parts, the equation of motion for
$x_j^1(t)$ which gives the equations of motion for the
bath variables $x_j (t)$ (from Eq.(\ref{7})) as
\begin{eqnarray}
x_j (t) = & \left [ x_j^0 (0) -c_j f(q(0)) \right ] \cos \omega_j t 
+ \frac{p_j^0 (0)}{\omega_j} \sin \omega_j t 
\nonumber \\
& + c_j \int_0^t dt' \cos \omega_j (t-t') f'(q(t')) \dot{q} (t')
\nonumber \\
& - \frac{\kappa_j}{\omega_j} 
\int_0^t dt' \sin \omega_j (t-t') \epsilon (t') x_j^0 (t').
\label{neq2}
\end{eqnarray}

\noindent
Using the above solution in Eq.(\ref{5}) we finally obtain the
equation of motion for system variable as
\begin{eqnarray}
\ddot{q} (t) = & -V'(q(t)) \nonumber \\
& + f'(q(t)) \sum_j c_j \omega_j^2
\left [ \left \{ x_j^0 (0) - c_j f(q(0)) \right \} \cos \omega_j t
+ \frac{p_j^0 (0)}{\omega_j} \sin \omega_j t \right ]
\nonumber \\
& - \sum_j c_j^2 \omega_j^2 f'(q(t)) \int_0^t dt' \cos \omega_j (t-t')
f'(q(t')) p(t')
\nonumber \\
& - f'(q(t)) \sum_j c_j \kappa_j \omega_j \int_0^t dt' \sin \omega_j (t-t')
\epsilon (t') x_j^0 (t'), \label{neq3}
\end{eqnarray}

\noindent
where $p(t)=\dot{q}(t)$ is the generalized momentum of the system variable.
This equation can be rewritten as
\begin{eqnarray} \label{12}
\ddot{q} (t) =  & -V'(q(t)) - f'(q(t))\int_0^t dt' \gamma(t-t')f'[q(t')]p(t') 
+ f'(q(t))F(t) \nonumber \\
& - f'(q(t)) \sum_{j}^{N} c_j\kappa_j\omega_j \int_0^t
dt' \sin\omega_j(t-t')x_j^0(t')\epsilon(t'),
\end{eqnarray}

\noindent where we have defined $\gamma (t)$ and $F(t)$ as,
$\gamma(t)=\sum_{j=1}^{N} c_j^{2} \omega_{j}^2 \cos
\omega_j(t)$ and $F(t) = \sum_{j=1}^{N} c_j \omega_{j}^2 [\{x_{j}(0)
- c_j f(q(0)\}\cos \omega_j t + (p_j(0)/\omega_j) \sin \omega_j t
]$. At this point, we note that, the forcing term $F(t)$ is
deterministic as expected. It ceases to be deterministic when it is not
possible to specify all the $x_j^0(0)$-s and $p_j^0(0)$-s, i.e., the 
initial conditions of all the bath variables, exactly. The
standard procedure to overcome this problem is to consider a distribution
of $x_j^0 (0)$ and $p_j^0 (0)$ to specify the statistical properties of the
bath dependent forcing term $F(t)$. The distribution of the bath oscillators
is assumed to be a canonical distribution of Gaussian form
\begin{equation}
W \left [ x_j^0 (0), p_j^0 (0) \right ]
= Z^{-1} \exp \left [ -\frac{H_B + H_{SB}}{k_BT} \right ]
\label{neq4}
\end{equation}

\noindent
where $Z$ is the bath partition function. This choice of the distribution
function of bath variables makes the internal noise $F(t)$ Gaussian.
It is now easy to verify the statistical properties of $F(t)$ as
$\langle F (t)
\rangle = 0$ and $\langle F(t) F(t')\rangle = 2 k_B T \gamma
(t-t')$. where $k_B$ is the Boltzmann constant and $T$ is the
equilibrium temperature. $\langle \cdot \rangle$ implies the
average over the initial distributions of bath variables which is
assumed to be a canonical distribution of Gaussian form as given in
Eq.(\ref{neq4}). The second relation is the celebrated fluctuation-dissipation
relation \cite{Kubo}
which ensures that the bath was in thermal equilibrium at $t=0$.

To identify Eq.(\ref{12}) as a generalized Langevin equation we must impose
some conditions on the coupling coefficients $c_j$ and $\kappa_j$, on the
bath frequencies $\omega_j$ and on the number $N$ of the bath oscillators
that will ensure that $\gamma (t)$ is indeed dissipative and the last term
in Eq.(\ref{12}) is finite for $N\rightarrow \infty$. A sufficient condition
for $\gamma (t)$ to be dissipative is that it is positive definite and 
decreases monotonically with time. These conditions are achieved if
$N \rightarrow \infty$ and if $c_j \omega_j^2$ and $\omega_j$ are
sufficiently smooth function of $j$ \cite{Ford}. As $N \rightarrow \infty$
one replaces the sum by an integral over $\omega$ weighted by a density
of state $\rho (\omega)$.
Thus to obtain a finite result in the continuum limit the coupling
function $c_i=c(\omega)$ and $\kappa_i=\kappa(\omega)$ are chosen
\cite{JRC,Bravo} as $c(\omega)= c_0 / \omega\sqrt\tau_c$ and
$\kappa(\omega)=\kappa_0$, where $c_0$ and $\kappa_0$ are constants
and $\tau_c$ is the correlation time of the heat bath. The choice
$\kappa (\omega)=\kappa_0$ is the simplest one where we assume that
every bath mode is excited with the same intensity. This simple choice
makes the relevant term finite for $N \rightarrow \infty$.
Consequently
$\gamma(t)$ becomes, $\gamma(t) = (c_0^2/\tau_c) \int d\omega
\rho(\omega) \cos \omega t$. $1/\tau_c$ may be characterized as
the cutoff frequency of the bath oscillators. The density of modes
$\rho(\omega)$ of the heat bath is assumed to be Lorentzian,
$\rho(\omega)=(2/\pi)[\tau_c/(1+\omega^2\tau_c^2)]$. The above
assumption resembles broadly the behavior of the hydrodynamical
modes in a macroscopic system \cite{Resibois}. With these forms of
$\rho(\omega)$, $c(\omega)$ and $\kappa(\omega)$ we have the
expression for $\gamma(t)$ as $\gamma(t) = (c_0^2 / \tau_c)
\exp(-t/\tau_c)$, which reduces to $\gamma(t)=2 c_0^2\delta(t)$ for
vanishingly small correlation time $\tau_c$ and consequently one obtains a
$\delta$-correlated noise process.


Taking into consideration of all the above assumptions and assuming
that the system variable evolves much more slowly in comparison to
the external noise $\epsilon(t)$, in the limit $\tau_c\rightarrow 0$,
Eq.(\ref{12}) reduces to
\begin{equation} \label{19}
\fl
\ddot{q} (t) = -V'(q(t)) - \gamma[f'(q(t))]^2 \dot{q}(t) + f'(q(t)) F(t) + \gamma
\kappa_0 f(q(t))f'(q(t)) \epsilon (t),
\end{equation}

\noindent where $\gamma = c_0^2$ is the dissipation constant and the
Langevin force $F(t)$ is characterized by the statistical properties
\begin{equation}\label{20}
\langle F (t) \rangle = 0, \langle F(t) F(t')\rangle =
2\gamma k_B T \delta (t-t').
\end{equation}

\noindent
Eq.(\ref{19}) is the generalized Langevin equation for the system variable.
At this juncture, it is noteworthy that for $f(q)=q$, Eq.(\ref{19}) reduces to
$\ddot{q} (t) = -V^{\prime} (q(t)) - \gamma \dot{q} (t) + F(t) 
+ \gamma \kappa_0 q (t)\epsilon(t)$.
Thus for linear system-bath coupling (i.e. for $f(q)=q$),
our Hamiltonian given by Eq.(\ref{1}) may be the starting point for the 
construction of a Langevin equation with both additive and multiplicative noise, 
which has numerous applications in various field of physics, e.g. phase transition etc.
For harmonic potential, this equation has been extensively studied by many authors
in various contexts \cite{lindenberg}. 

In the above Langevin equation (\ref{19}) the noise terms (internal and
external) appear multiplicatively and the dissipation is space dependent.
Using the method of van Kampen \cite{lindenberg} for nonlinear stochastic
differential equations the Fokker-Planck equation corresponding to the
Langevin equation (\ref{19}) is given by \cite{JRC,lindenberg}
\begin{eqnarray}
\frac{\partial P}{\partial t} = & -\frac{\partial}{\partial q} (pP)
+ \frac{\partial}{\partial p} \left \{ \lambda(q) p + V' (q) \right \} P
\nonumber \\
& + \left \{ \lambda (q) k_BT + \gamma^2 \kappa_0^2 D \left [
f(q) f'(q) \right ]^2 \right \} \frac{\partial^2 P}{\partial p^2}
\label{neq5}
\end{eqnarray}

\noindent
where $P = P(q,p,t)$ is the phase space probability density function and 
$\lambda(q)=\gamma [f'(q)]^2$ is the space dependent dissipation function.
Instead of handling two noise processes (internal and external) independently,
one can define an \textit{effective} noise process $\xi (t)$ and an
auxiliary function $G(q)$ to obtain the same Fokker-Planck equation 
(\ref{neq5}) from the following Langevin equation
\begin{equation}\label{21}
\ddot{q} = -V'(q) - \lambda(q) \dot{q} + G(q)\xi(t),
\end{equation}

\noindent with
\begin{eqnarray}\label{22}
\langle \xi (t) \rangle & = 0, \langle \xi(t) \xi(t')\rangle
= 2\delta (t-t'), \\
\lambda(q)& = \gamma [f'(q)]^2, G(q) = f'(q)\sqrt{\gamma k_B T +
D(\gamma \kappa_0)^2 f^2(q)}.
\end{eqnarray}

\noindent That the Langevin equation (\ref{21}) gives the same Fokker-Planck
equation (\ref{neq5}) can be verified by using van Kampen's
methodology \cite{lindenberg}. The construction of Langevin
equation using an effective noise term and an auxiliary function has been
done earlier in the configuration space by Wu \textit{et al} \cite{Wu},
whereas we have written the Langevin equation (\ref{21}) in the phase space.
Thus as far as the equation for the evolution of probability density function
is concerned, Eq.(\ref{21}) is the equivalent description of the stochastic
differential equation (\ref{19}). Eq.(\ref{21}) is one of the 
\emph{key results} of this work as it incorporates the effects of thermal noise
$F(t)$ and the external noise $\epsilon (t)$ in an unified way even when
the underlying noise processes are multiplicative due to the nonlinear
system-bath coupling and nonlinear modulation of the heat bath by an external
noise. It is important to mention here that Eq.(\ref{21}) describes a 
thermodynamically open system where there is no fluctuation-dissipation
relation so that the system will not reach at usual thermal equilibrium,
instead, a steady state is attainable for large $t$ \cite{JRC,lindenberg}.
From the computational point of view, generation of a single multiplicative
noise process is much more economical than to generate two separate 
multiplicative noise processes.

In Eq.(\ref{21}) the noise is multiplicative and the
dissipation is space dependent. In the case of large dissipation one
eliminates the fast variables adiabatically to get a simpler
description of the system dynamics. The traditional approach to the
elimination of fast variables for multiplicative noise processes
does not always give the correct description. In order to get the
correct Langevin equation in the overdamped limit we follow the
method of Sancho \textit{et al} \cite{Sancho} and then using van
Kampen's lemma \cite{NGVK} and Novikov's theorem \cite{Novikov} we
get the Fokker-Planck-Smoluchowski equation
corresponding to Eq.(\ref{21}) for the probability density $P(q,t)$
in the configuration space \cite{Sancho}
\begin{equation}\label{23}
\frac{\partial P(q,t)}{\partial t}=\frac{\partial}{\partial
q}\frac{1}{\lambda(q)}\left[V'(q)+\frac{\partial}{\partial
q}\frac{G^2(q)}{\lambda(q)}\right]P(q,t) .
\end{equation}

\noindent
In the ordinary Stratonovich description \cite{Sancho} the Langevin
equation corresponding to the Fokker-Planck equation (\ref{23}) is
\begin{equation}
\label{neq6}
\dot{q} = -\frac{V'(q)}{\lambda (q)} - 
\frac{G(q) G'(q)}{[\lambda (q)]^2} + \frac{G(q)}{\lambda (q)} \xi (t).
\end{equation}

\noindent
The above equation (\ref{neq6}) differs from the Langevin equation, obtained 
by using the traditional way of adiabatic elimination of fast variable, due to
the presence of the second term on the right hand side. This term,
$G(q)G'(q)/[\lambda (q)]^2$, represents the effect of multiplicative noise
in the process of elimination of fast variable \cite{Sancho}.

The stationary solution of Eq.(\ref{23}) contains
inhomogeneous effective temperature like term, a generalization of
Boltzmann factor for state dependent diffusion in open system, which
arises due to the entanglement of the external driving with the
nonlinearity of the system heat bath coupling, a well known effect
in several contexts, \textit{e.g.}, Landauer Blow torch effect
\cite{Landauer}. In the absence of external bath modulation, i.e., when
$G(q)=f'(q)\sqrt{\gamma k_BT}$, Eq.(\ref{23}) gives the correct equilibrium
distribution function, $P_{eq}(q)={\cal N} \exp [-V(q)/k_BT]$ with
${\cal N}$ being the normalization constant.
In the overdamped limit we then have the stationary current as
\begin{equation}\label{24}
J=-\frac{1}{\lambda(q)}\left[V'(q)+\frac{d}{dq}\left(\frac{G^2(q)}{\lambda(q)}\right)
\right]P_{st}(q) .
\end{equation}

\noindent Integrating the above equation we have the expression of
stationary probability distribution in terms of stationary current
\begin{equation}\label{25}
P_{st}(q) =\frac{e^{-\phi(q)}}{G^2(q)/\lambda(q)}
\left[\frac{G^2(0)}{\lambda(0)}P_{st}(0) -J\int_0^q
\lambda(q')e^{\phi(q')}dq'\right]
\end{equation}

\noindent
where
\begin{equation} \label{26}
\phi(q) =\int_0^q \frac{V'(q')}{G^2(q')/\lambda(q')} dq' = \int_0^q
\frac{V'(q')}{k_B T + D \gamma \kappa_0^2 f^2(q')} dq'.
\end{equation}

\noindent We then consider a symmetric periodic potential with
periodicity $2\pi$, $V(q)=V(q+2\pi)$ and the periodic coupling
function with the same periodicity as the potential,
$f(q)=f(q+2\pi)$. Now applying the periodic boundary condition on
$P_{st}(q)$, $P_{st}(q)=P_{st}(q+2\pi)$ and the normalization
condition on stationary probability distribution we have the
expression for stationary current \cite{Risken}
\begin{eqnarray}\label{30}
J & =\left[1-e^{\phi(2\pi)}\right]/\left\{ \int_0^{2\pi}
\frac{\lambda(q)}{G^2(q)}e^{-\phi(q)} dq \int_0^{2\pi}
\lambda(q')e^{\phi(q')} dq' -
\left[1-e^{\phi(2\pi)}\right] \right. \nonumber \\
& \left. \times \int_0^{2\pi} \frac{\lambda(q)}{G^2(q)}e^{-\phi(q)}
\int_0^{q} \lambda(q')e^{\phi(q')} dq' dq \right\}.
\end{eqnarray}

\noindent From the condition of periodicity it is clear that for the
periodic potential and the periodic derivative of coupling function
with same periodicity $V'(q)/[G^2(q)/\lambda(q)]$ is periodic with
same periodicity. This makes the effective potential $\phi(2\pi)$
equal to zero so that the numerator of Eq.(\ref{30}) reduces to
zero. Thus there is no occurrence of current for a periodic
potential and periodic derivative of coupling function with same
periodicity and hence there is no violation of second law of
thermodynamics. The thermodynamic consistency based on symmetry
consideration ensures the validity of the present formalism.
B\"{u}ttiker \cite{Buttiker} have shown that a overdamped particle
subjected to a drift force field with sinusoidal space dependence
and also a sinusoidally modulated space dependent diffusion with the
same period as the drift experiences a net driving force. The
resulting current depends on the amplitude of the modulation of
diffusion and is a periodic function of phase difference between the
sinusoidal drift and the sinusoidal modulation of the diffusion.

\begin{figure}[t]
\includegraphics{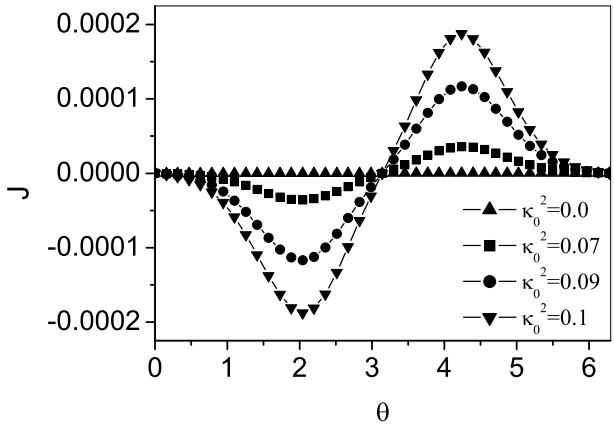}
\caption{\label{fig1}  Variation of current, $J$ as a function of
phase difference, $\theta$ for different values $\kappa_0^2$ and for
the parameter set $\alpha=0.5$, $k_BT=0.1$, $\gamma=1.0$ and
$D=1.0$.}
\end{figure}

Let us consider that the particle is moving in a sinusoidal
symmetric potential of the form
\begin{equation}\label{31}
V(q)=V_0[1+\cos(q+\theta)],
\end{equation}

\noindent where $V_0$ is constant and may be taken as barrier height
and $\theta$ is the phase factor which can be controlled externally.
The coupling function  is chosen as $f(q)=q + \alpha \cos q$, where
$\alpha$ is the modulation parameter. We now calculate the current
given by Eq.(\ref{30}). In Fig.1 the variation of current as a
function of phase difference is shown for different values of the
coupling constant $\kappa_0$. Since $\kappa_0$ is the perturbation
parameter in our analysis we have kept its maximum value low, i.e.,
$\sim k_BT$. For the value of the other parameters we have chosen a
particular set from the complete parameter space. An extensive
analysis using the full parameter space will be given in our future
communication. The current shown in Fig.(1) is basically due to the
phase difference between the symmetric periodic potential and the
space dependent diffusion caused by the nonlinear modulation of the
heat bath by external noise. The current does vanish when the phase
difference is either zero or integral multiple of $\pi$. When the
heat bath is linearly modulated by external noise source, it is easy
to observe that the effective potential $\phi(q)$ is integrable and
there will be no asymmetry in the effective potential as the noise
in the corresponding Langevin equation appears additively and the
diffusion coefficient becomes space independent. Thus when the heat
bath is driven nonlinearly by the external noise agency there is a
net directed motion or phase induced current. This is because of the
fact that when the external noise drives the heat bath nonlinearly
the phase bias gives a tilt to the effective potential $\phi(q)$
which makes the transition between left to right and right to left
unequal. In Fig.2 we plot the generalized potential $\phi(q)$ for
various coupling constant $\kappa_0^2$. The phase difference (hence
the nonlinear driving of the heat bath) breaks the detailed balance
of the system. When the phase difference is zero or the heat bath is
driven linearly there is no net drift velocity. Thus, when we drive
the heat bath linearly with $\delta$-correlated external noise, even
in presence of phase difference between $V(q)$ and $f'(q)$ there is
no net current. For net drift, apart from phase difference nonlinear
driving of the heat bath is required. This is the \textit{central
result of this paper}.

In conclusion, we have proposed a new microscopic analysis to study
the generation of directed motion for a nonlinearly driven heat bath
by an external noise. Making use of a perturbative treatment we have
derived an effective Langevin equation with space dependent
dissipation and multiplicative noise. Using the corresponding
Fokker-Planck-Smoluchowski equation with a space dependent diffusion
coefficient we have checked the thermodynamic consistency condition
and have shown that to observe a phase induced current one
necessarily needs a nonlinear driving of the heat bath by an
external Gaussian noise. In our future venture in this direction we
wish to compare our analytical result with stochastic simulation
using the complete parameter space.

\begin{figure}[t]
\includegraphics{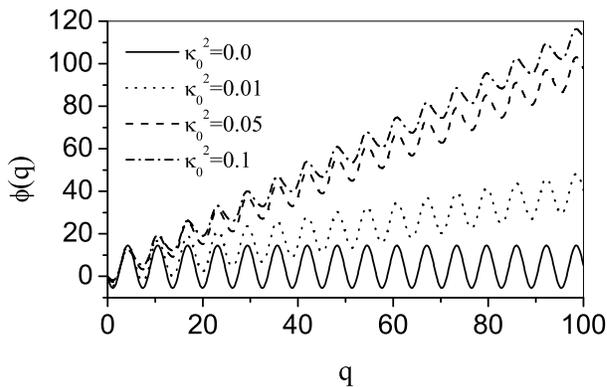}
\caption{\label{fig2} Plot of the generalized potential $\phi(q)$ as
a function of coordinate $q$ for different values of $\kappa_0^2$
and for the parameter set $\alpha=0.5$, $k_BT=0.1$, $\gamma=1.0$,
$D=1.0$ and $\theta=0.5\pi$.}
\end{figure}

We specially thank Professor J. K. Bhattacharjee for constructive
suggestions. Stimulating discussions with Dr. B. Deb, Professor Eli
Pollak, and Professor D. S. Ray is thankfully acknowledged. JRC is
thankful to Indian Academy of Sciences, Bangalore for financial
support. SKB acknowledges support from Virginia Tech through ASPIRES
award program.

\end{document}